\documentclass[12pt,a4paper]{article}
\pdfoutput=1
\usepackage{cite}
\usepackage[a4paper,total={6in, 9in}]{geometry}
\usepackage{graphicx}
\usepackage{dcolumn}
\usepackage{bm}
\usepackage{amsmath}
\usepackage{amssymb}
\usepackage{float}
\usepackage{appendix}
\usepackage{url}
\usepackage[normalem]{ulem}
\usepackage{xcolor}
\usepackage{listings}
\newcommand{\drv}{{\rm d}}

\title{\bf Forward Drell--Yan production at the LHC in the BFKL formalism with collinear corrections}
\author{
F. G. Celiberto, D. Gordo G\'omez, A. Sabio Vera \\ \\
{
\small Instituto de F{\' \i}sica Te{\' o}rica UAM/CSIC, Nicol{\'a}s Cabrera 15} \\ 
{\small \& Universidad Aut{\' o}noma de Madrid, E-28049 Madrid, Spain.}
}

\begin{document}

\maketitle 

\abstract

Motivated by the recent work of Brzemi{\' n}ski, Motyka, Sadzikowski and Stebel in~\cite{Brzeminski:2016lwh}, where forward Drell--Yan production is studied in proton-proton collisions at the LHC, we improve their calculation by introducing  an unintegrated gluon density obtained in~\cite{HERAfitH_SV_S} from a fit to combined HERA data at small values of Bjorken $x$. This gluon density was calculated within the BFKL formalism at next-to-leading order with collinear corrections. We show that it generates a good description of the forward Drell--Yan cross section dependence on the invariant mass of the lepton pair both for LHCb and ATLAS data.

\section{Introduction}

For many years forward  Drell--Yan (DY) production at hadron colliders~\cite{DYbunch} has been proposed as an interesting observable to prove the unintegrated gluon density in the proton at very small values of Bjorken-$x$~\cite{Anderson:09/10jca,Jochen}. 
In the recent work of Brzemi{\' n}ski, Motyka, Sadzikowski and Stebel in~\cite{Brzeminski:2016lwh} this observable has been studied in the context of high energy resummations at small $x$. They investigated this process in proton-proton scattering at the LHC with $\sqrt{s}=14$ TeV. Their focus lied on the dipole formalism with saturation corrections, using the Golec-Biernat--W{\" u}sthoff model~\cite{GolecBiernat:1998js}, or without them, making use of the Balitsky--Fadin--Kuraev--Lipatov (BFKL) approach at leading order (LO). Their investigations centered around the idea of what is the different twist content in each approach (see also their previous related work in~\cite{Motyka:2014lya}). 

In the present letter we shift the focus showing that it is possible to obtain a good description of the latest DY data for small values  of the lepton pair masses when an unintegrated gluon density calculated within the next-to-leading order (NLO) BFKL formalism with collinear corrections is used. This shows that it is possible to obtain a correct description of HERA structure functions $F_2$ and $F_L$ together with LHC DY data using a common approach based on the NLO BFKL formalism. Our calculation is particularly relevant in the low DY pair mass region, where we focus the discussion. The results here presented should also be compared to those in~\cite{GolecBiernat:2010de,Ducati:2013cga,Schafer:2016qmk,Ducloue:2017zfd}.

\begin{figure}[tb]
\centering
 \includegraphics[scale=0.40,clip]{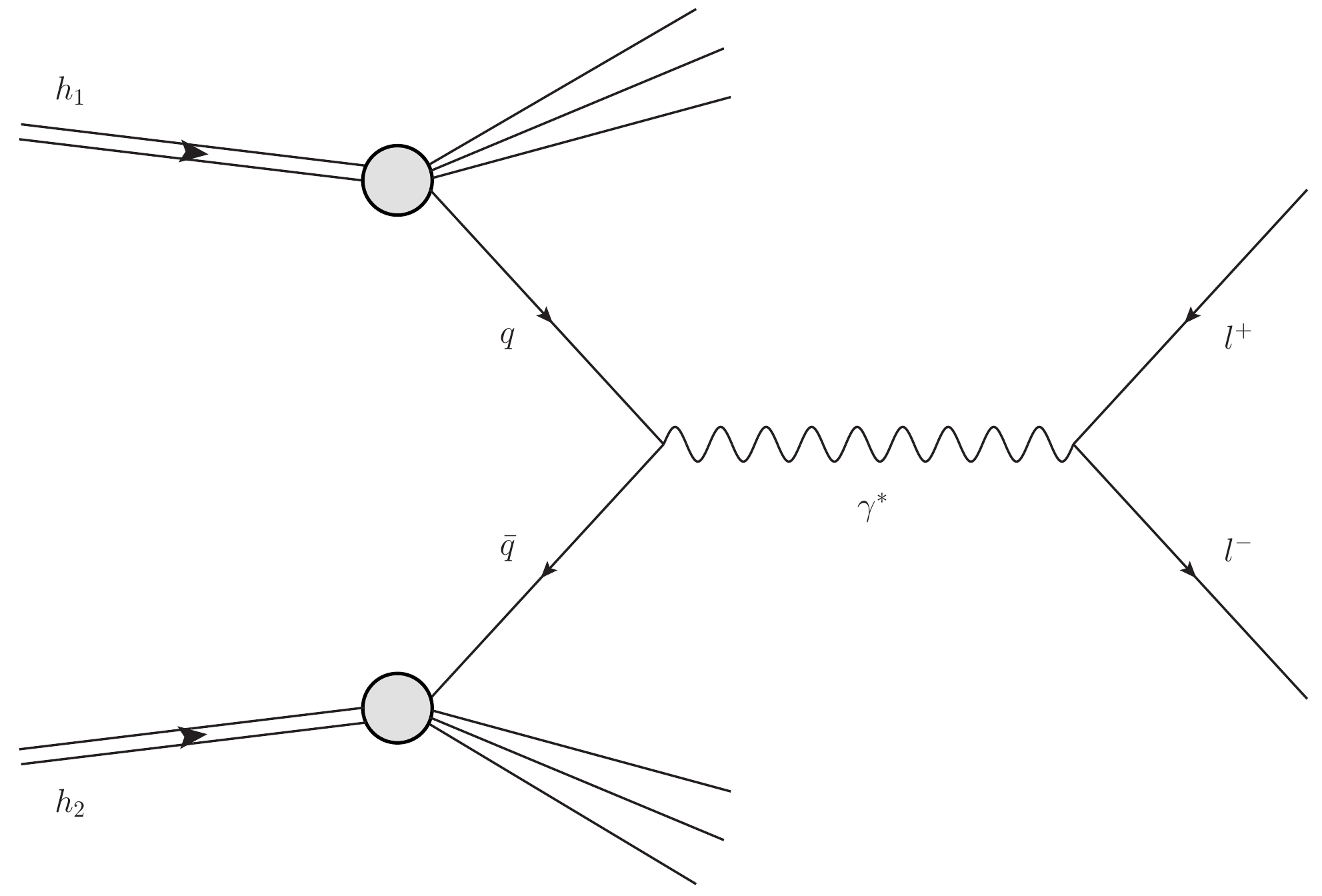}
\caption{Drell--Yan dilepton production at leading order.}
\label{DYLO}
\end{figure}

To be more precise, we study the production of a lepton-antilepton pair, $L^+ L^-$, in proton-proton ($p p$) collisions as shown in Fig.~\ref{DYLO} with notation
\begin{equation}
\label{eq:proc}
 p(P_1) \, + \, p(P_2) \; \to \; L^+(l^+) \, + \, L^-(l^-) \, + \, X
\end{equation}
where $X$ indicates any inclusive secondary radiation. At leading order in the electroweak coupling  this process is mediated by a virtual photon $\gamma^* (q)$ or a $Z^0 (q)$ boson, where the vector boson momentum $q = l^+ + l^-$ carries a $q_T$ transverse component. $M^2 \equiv q^2 > 0$ corresponds to the lepton pair invariant mass squared.
The angular distributions of the lepton pair can be expressed in terms of four independent structure functions, $\cal{W}_{[\lambda]}$. 
The DY standard formulation~\cite{Lam:1978pu} allows for a factorization between the lepton pair angular phase space $\drv\vartheta^* \, \drv\varphi^* \equiv \drv\Omega_l^*$, where $\vartheta^*$ and $\varphi^*$ are the polar and azimuthal angles of the lepton momentum vector in the dilepton center-of-mass frame, and the structure functions, defined as projections of the DY amplitudes on the exchanged boson. 

In the present work we neglect the $Z^0$ contribution which is only relevant at higher $M^2$ (we will also show the experimental data points at the highest values of $M^2$ just to gauge the importance of the missing diagrams). The DY differential cross section hence reads
\begin{equation}
\label{eq:dsigma}
 \frac{\drv \sigma}{\drv M \, \drv \Omega_l^* \, \drv x_F \, \drv q_T}
 = \frac{\alpha^2 \, q_T}{(2 \pi \, M)^3}
 \left[
 \left( 1 - \cos^2 \vartheta^* \right) {\cal W}_L
 + \left( 1 + \cos^2 \vartheta^* \right) {\cal W}_T
 \right.
\end{equation}
\[
 \left.
 + \; \left( \sin 2 \vartheta^* \cos \varphi^* \right) {\cal W}_\Delta
 + \left( \sin^2 \vartheta^* \cos 2 \varphi^* \right) {\cal W}_{\Delta \Delta}
 \right]
 \; ,
\]
where $x_F$ stands for the Feynman variable representing the longitudinal momentum fraction from the initial-state hadron carried by the virtual photon, $\alpha$ is the electromagnetic coupling constant, ${\cal W}_L$ and ${\cal W}_T$ are structure functions for longitudinally and transversely polarized virtual photons, respectively, ${\cal W}_\Delta$ is the single-spin-flip structure function, and ${\cal W}_{\Delta \Delta}$ is the double-spin-flip one. The frame orientation is not unique and determines the form of the structure functions. In this work the choice of the Gottfried--Jackson frame~\cite{Gottfried:1964nx}, with the Z axis anti-parallel to the target's momentum and the Y axis orthogonal to the reaction plane, is employed.

In the high energy factorization approach, it is possible to write  the structure functions as the convolution
\begin{equation}
\label{eq:W-HEF}
 {\cal W}_{[\lambda]}\hspace{-.1cm} = \hspace{-.1cm}\frac{2 \pi}{3} \alpha_s (\mu_R) M^2 \hspace{-.2cm} \int_{x_F}^1 \hspace{-.08cm}
 \frac{\drv z}{z^2} 
 f^* \left( \frac{x_F}{z}, \mu_F \right)\hspace{-.2cm}
 \int \hspace{-.15cm} \frac{\drv \kappa_T  \drv \phi_{\kappa_T}}{\left( \kappa_T^2 \right)^2}  \, {\cal G} (x_g, \kappa_T^2) \, \Phi_{[\lambda]} (q_T, \vec{\kappa}_T, z).
\end{equation}
\begin{figure}
\centering
 \includegraphics[scale=0.32849,clip]{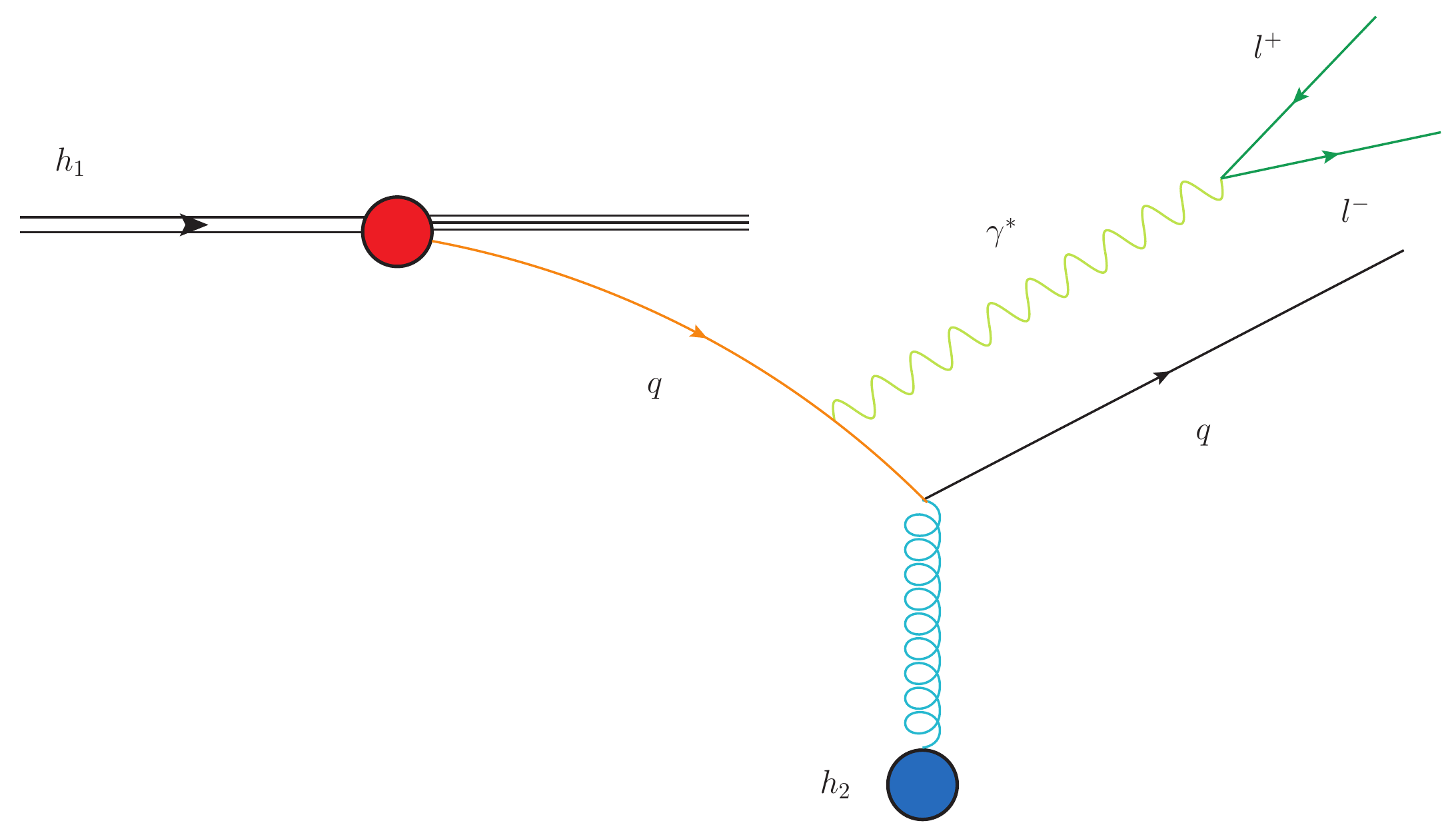}
 \includegraphics[scale=0.32849,clip]{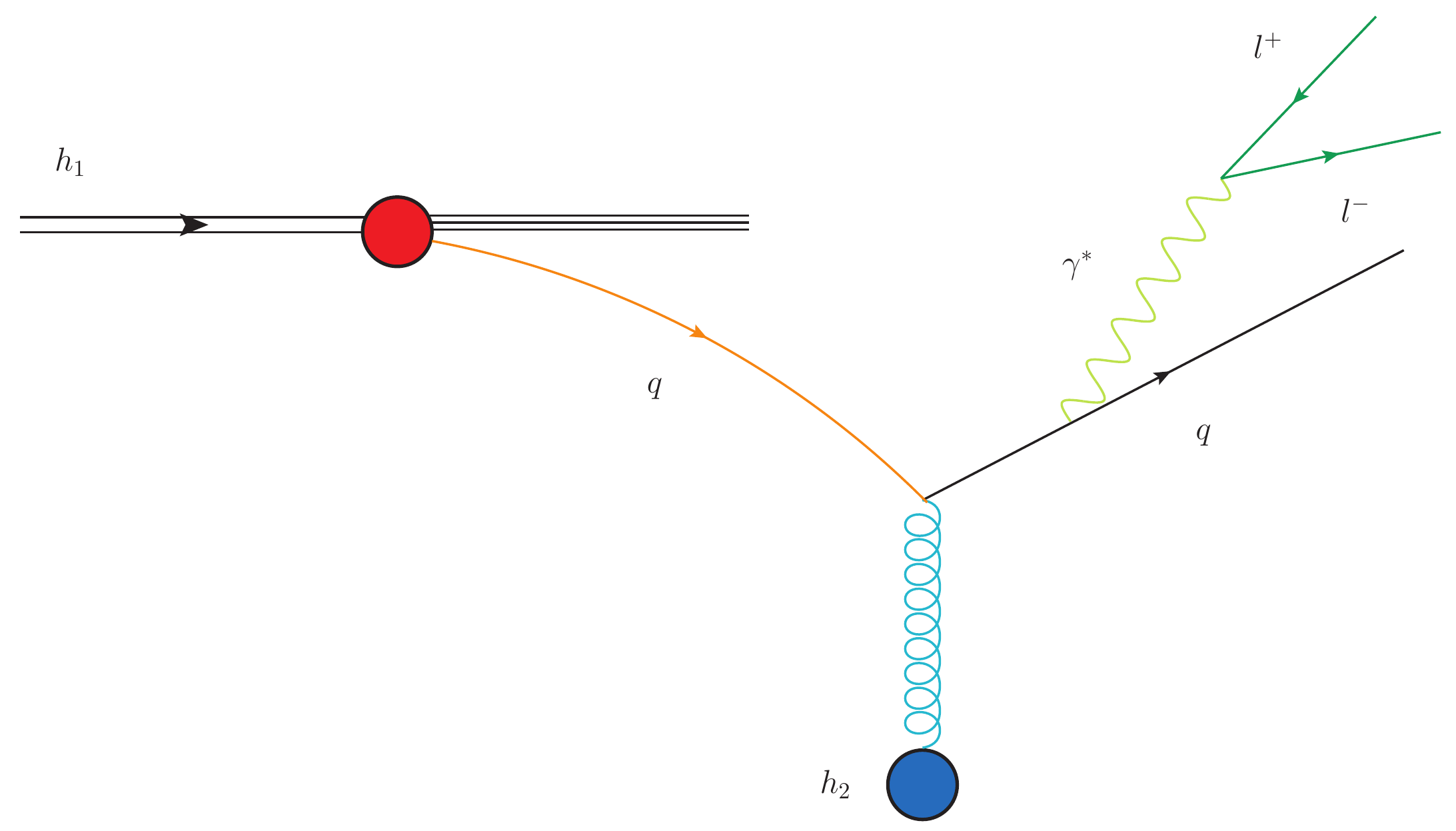}
\caption{Compton scattering diagrams for DY dilepton forward production.}
\label{fig:DY-foward-NLO}
\end{figure}
\hspace{-.1cm}$z$ is the longitudinal momentum fraction of the initial-state quark carried by the virtual photon, $\kappa_T \equiv |\vec{\kappa_T}|$ and  $q_T \equiv |\vec{q_T}|$. We use a collinear parton distribution function, $f^* \left( x, \mu_F \right) = \sum_{r} f_r \left( x, \mu_F \right)$ which accounts for the incident u, d, s, c and b quarks and corresponding antiquarks with a high-$x$ value (orange lines in Fig.~\ref{fig:DY-foward-NLO}). 

We also have an unintegrated, transverse momentum dependent, gluon distribution function, ${\cal G} (x_g, \kappa_T^2)$, carrying all the information about the small-$x$ gluon evolution (blue lines in Fig.~\ref{fig:DY-foward-NLO}), and the forward DY impact factors, $\Phi_{[\lambda]} (q_T, \vec{\kappa}_T, z)$, accounting for the $\gamma^* \, \to \, L^+ L^-$ transition. The gluon longitudinal momentum fraction $x_g$ follows from the forward DY kinematics in the $qg^* \, \to \, q\gamma^*$ channel, {\it i.e.} (with $s = (P_1 + P_2)^2$ being the  center-of-mass energy squared)
\begin{equation}
\label{eq:x_g}
 x_g = \frac{M^2 (1 - z) + q_T^2 + z(\kappa_T^2 - 2 \vec{\kappa}_T \cdot \vec{q}_T )}{s \, x_F \, (1 - z)} 
 \approx \frac{M^2 (1 - z) + q_T^2}{s \, x_F \, (1 - z)}.
\end{equation}

After this brief Introduction regarding the process of interest and our calculational set up, we now give some details related to the unintegrated gluon density used in our calculations and the structure of the forward impact factors. We then present our results and Conclusions. 

\section{Unintegrated gluon distribution and forward impact factors}

The standard definition of the small-$x$ transverse momentum dependent gluon distribution, better known as unintegrated gluon distribution, relies on the convolution between the universal BFKL gluon Green's function, which takes into account the resummation of high-energy logarithms, and the proton impact factor, which describes the coupling of the gluon Green's function to the proton. This proton impact factor is characterized by large transverse scales and therefore, being of non-perturbative nature, needs to be modeled. 

We use the three-parameter model for the coupling of the gluon Green's function and the proton 
put forward in the study of deep inelastic scattering structure functions 
in~\cite{HERAfitH_SV_S}. More recently, it has been used to investigate single-bottom quark production at the LHC~\cite{Chachamis:2015ona},
$J/\Psi$ and $\Upsilon$ photoprotoduction~\cite{Bautista:2016xnp} and $\rho$-meson leptoproduction at HERA~\cite{Bolognino:2018rhb}. In transverse momentum space, it simply reads
\begin{equation}
\label{eq:proton_if}
 \Phi_p(q, Q_0^2) = \frac{{\cal C}}{2\pi \Gamma(\delta)}
 \left(\frac{q^2}{Q_0^2}\right)^\delta e^{-\frac{q^2}{Q_0^2}}
\end{equation}
having a maximum at $p^2 = \delta Q^2_0$. 
The values of the parameters $Q_0$ = 0.28 GeV, $\delta$ = 8.4 and ${\cal C}$ = 1.50, were obtained from a fit to combined HERA data~\cite{HERAfitH_SV_S} when the leading order photon impact factor was used. Since we are also using a leading order calculation for the vertex producing the DY pair, it is consistent to use the same set of parameters in our current analysis. 

Combining Eq.~(\ref{eq:proton_if}) with the gluon Green's function, we obtain the following expression for the unintegrated gluon density:
\begin{equation}
\label{eq:UGD}
 {\cal G}(x, \kappa^2, \mu_H) = \int_{-\infty}^{\infty}
  \frac{d\nu}{2\pi^2}\ {\cal C} \  \frac{\Gamma(\delta - i\nu -\frac{1}{2})}
  {\Gamma(\delta)}\ \left(\frac{1}{x}\right)^{\chi\left(\frac{1}{2}+i\nu\right)}
  \left(\frac{\kappa^2}{Q^2_0}\right)^{\frac{1}{2}+i\nu}
\end{equation}
\[
\times \left\{ 1 +\frac{\bar{\alpha}^2_s \beta_0 \chi_0\left(\frac{1}{2}
+i\nu\right)}{8 N_c}\log\left(\frac{1}{x}\right)
\left[-\psi\left(\delta-\frac{1}{2} - i\nu\right)
-\log\frac{\kappa^2}{\mu_H^2}\right]\right\}\,,
\]
with $\beta_0=\frac{11 N_c-2 N_f}{3}$  the first coefficient of the QCD $\beta$-function, $N_f = 5$, $\bar{\alpha}_s \equiv N_c / \pi \alpha_s\left(\mu^2\right)$, and $\mu^2 = \mu_H Q_0$.
$ \chi_0(\gamma) = 2\psi(1) - \psi(\gamma) - \psi(1-\gamma)$, with $\gamma \equiv \frac{1}{2} + i\nu$, is the LO eigenvalue of the BFKL
kernel and $\psi(z) = \Gamma^\prime(z)/\Gamma(z)$. $\mu_H$ is a characteristic hard scale which can be set equal to the photon invariant mass, $M$. Finally, $\chi(\gamma)$ is the NLO eigenvalue of the BFKL kernel, 
\begin{equation}
\label{eq:chi}
 \chi(\gamma) = \bar{\alpha}_s\chi_0(\gamma)+\bar{\alpha}^2_s\chi_1(\gamma) -\frac{1}{2}\bar{\alpha}^2_s\chi^\prime_0(\gamma)\,\chi_0(\gamma) + \chi_{RG}(\bar{\alpha}_s, \gamma)\,,
\end{equation}
with $\chi_1(\gamma)$ and $\chi_{RG}(\bar{\alpha}_s, \gamma)$ (which includes the collinear corrections resummed in the form of a Bessel function as calculated in~\cite{Vera:2005jt}) given in
Section~2 of Ref.~\cite{Chachamis:2015ona}, to which we refer for further details (also on the particular treatment of the running of the coupling).

For completeness, we now briefly write down the expressions for the forward dilepton impact factors used in our work. In $\kappa_T$-representation they can be computed combining Eq.~(3.5) with Eqs.~(3.12) and (3.13) of Ref~\cite{Motyka:2014lya}, and applying the relations given in Eqs.~(3.24)-(3.27) of the same Reference, {\it i.e.}
\begin{equation}
\label{eq:if_L_kt}
 \Phi_L (q_T, \vec{\kappa}_T, z) = 
 \frac{2 M^2 (1-z)^2 z^2 \left((z \vec{\kappa}_T  - 2 \, \vec{q}_T) \cdot \vec{\kappa}_T\right)^2}{[M^2 (1-z) + q_T^2]^2 \, [M^2 (1-z) + (\vec{q}_T - z \vec{\kappa}_T)^2]^2}  \; ,
\end{equation}
\begin{equation}
\label{eq:if_T_kt}
 \Phi_T (q_T, \vec{\kappa}_T, z) = \frac{1 + (1-z)^2}{2} \left[ \frac{(q_T - z \kappa_x)^2 - z^2 \kappa_y^2}{[M^2 (1-z) + (\vec{q}_T - z \vec{\kappa}_T)^2]^2}  \right.
\end{equation}
\[
 \left. + \frac{q_T^2}{[M^2 (1-z) + q_T^2]^2} + \frac{2 \, q_T \, (z \kappa_x - q_T)}{[M^2 (1-z) + q_T^2] \, [M^2 (1-z) + (\vec{q}_T - z \vec{\kappa}_T)^2]} \right] \; ,
\]
\begin{equation}
\label{eq:if_LT_kt}
 \Phi_\Delta (q_T, \vec{\kappa}_T, z) =  \left(q_T (z \vec{\kappa}_T - 2 \vec{q}_T) \cdot \vec{\kappa}_T + 
 \kappa_x (M^2 (1-z)+ q_T^2)\right) 
\end{equation}
\[
 \times \; \frac{ M  \, (2-z) \, (1-z) \, z^2 \, (z \, \vec{\kappa}_T - 2 \, \vec{q}_T) \cdot \vec{\kappa}_T}{[M^2 (1-z) + q_T^2]^2 \, [M^2 (1-z) + (\vec{q}_T - z \vec{\kappa}_T)^2]^2} \; ,
\]
\begin{equation}
\label{eq:if_TT_kt}
 \Phi_{\Delta \Delta} (q_T, \vec{\kappa}_T, z) = (z-1) \left[ \frac{(q_T - z \kappa_x)^2 - z^2 \kappa_y^2}{[M^2 (1-z) + (\vec{q}_T - z \vec{\kappa}_T)^2]^2}  \right.
\end{equation}
\[
 \left. + \frac{q_T^2}{[M^2 (1-z) + q_T^2]^2} + \frac{2 \, q_T \, (z \kappa_x - q_T)}{[M^2 (1-z) + q_T^2] \, [M^2 (1-z) + (\vec{q}_T - z \vec{\kappa}_T)^2]} \right] \; ,
\]
where $\kappa_x \equiv \kappa_T \cos \phi_{\kappa_T}$, $\kappa_y \equiv \kappa_T \sin \phi_{\kappa_T}$ and $\vec{q}_T \cdot \vec{\kappa_T} \equiv q_T \kappa_T \cos \phi_{\kappa_T}$.

Before moving forward to presenting our results together with the LHC data, let us indicate that, for comparison, we will present and compare our results with a LO BFKL model defined within the color dipole approach~\cite{Nikolaev:1990ja,Brodsky:1996nj,Kopeliovich:2000fb,Kopeliovich:2001hf}, which has also been used in the work of Motyka et al in~\cite{Brzeminski:2016lwh}. Analogously to the formula given in Eq.~(\ref{eq:W-HEF}), it is possible to write expressions for the helicity structure functions as the convolution
\begin{equation}
\label{eq:W-CDA}
 {\cal W}_{[\lambda]} = \int_{x_F}^1 \drv z \, f^* \left( \frac{x_F}{z}, \mu_F \right) \int\limits_{\frac{1}{2} - i \infty}^{\frac{1}{2} + i \infty} \frac{\drv \gamma}{2 \pi i} \,\, \hat{\sigma} (\gamma) \, \left[ \frac{z^2 \hat{Q}_0^2}{M^2 (1-z)} \right]^\gamma \, \hat{\Phi}_{[\lambda]} (q_T, \gamma, z) \; ,
\end{equation}
where $f^* \left( \frac{x_F}{z}, \mu_F \right)$ is the collinear quark parton distribution functions defined right below Eq.~(\ref{eq:W-HEF}), $\hat{\sigma} (\gamma)$ is the dipole proton cross section calculated in Mellin space, $\hat{Q}_0$ is the scale transform parameter and $\hat{\Phi}_{[\lambda]} (q_T, \gamma, z)$ are the Mellin-transformed impact factors, originally calculated in Ref.~\cite{Motyka:2014lya} (see Eqs.~(3.32)-(3.35) of the same Reference for their analytic expressions). As a LO BFKL model for the dipole cross section we follow~\cite{Brzeminski:2016lwh} and use
\begin{equation}
\label{eq:dipole_BFKL_sigma}
 \hat{\sigma} (\gamma) \equiv \hat{\sigma}_{\rm LL} (\gamma) = - \hat{\sigma}_0 \, \Gamma (-\gamma) \, e^{\bar{\alpha}_s \chi_0 (\gamma) y} 	\, ,
\end{equation}
where $ y = \log \left( \frac{x_A}{x_g} \right)$ with $x_A = 0.1$ and $ x_g = \frac{M^2 (1 - z) + q_T^2}{s \, x_F \, (1 - z)}$ is the evolution length in rapidity. The relevant parameters are fixed by a fit to the deep inelastic scattering data~\cite{Motyka:2014jpa}: $\hat{Q}_0 = 0.51$ GeV, $\hat{\sigma}_0 = 17.04$ mb, $\bar{\alpha}_s = 0.087$, while running-coupling effects are neglected.

We now present the results of our calculations. 

\section{Results and comparison to data}

We are interested in the study of the dependence on the dilepton invariant mass $M$ of the forward Drell--Yan cross section (Eq.~(\ref{eq:dsigma})):
\begin{equation}
\label{eq:sigma}
 \frac{\drv \sigma (M)}{\drv M} = \int \drv \Omega_l^* \int \drv x_F \int \drv q_T \; \frac{\drv \sigma}{\drv M \, \drv \Omega_l^* \, \drv x_F \, \drv q_T} \; .
\end{equation}
We will also provide predictions for the total cross section averaged on bins in the $M$ variable. In order to match the kinematical cuts on the dilepton phase space used by the LHCb collaboration~\cite{LHCb:2012fja} it is needed to perform a Lorentz boost from the dilepton frame to the collision center-of-mass frame. The corresponding boost parameters are
\begin{equation}
\label{eq:boost_gamma}
 \gamma_{\rm B} = \frac{x_F \sqrt{s}}{2 M} \beta^+
 \; , \quad
 \vec{v}_{\rm B} = - \left( \frac{\vec{q}_T}{\gamma_{\rm B} M}, \frac{\beta^-}{\beta^+} \right)_{x,y,z}
\end{equation}
with $ \beta^\pm = 1 \pm \frac{M^2 + q_T^2}{s \, x_F^2}$.  We can now give expressions for the lepton momenta in the collision frame:
\begin{equation}
\label{eq:boost_l}
 l^{\pm} \equiv \lvert \vec{l}^\pm  \rvert = E^{\pm} = \frac{\gamma_B M}{2} \left( 1 \mp \vec{v}_B \cdot \vec{u}_{\Omega_l^*}  \right) \; ,
\end{equation}
\begin{equation}
\label{eq:boost_lz}
 l^{z,\pm} = \frac{M}{2} \left[  \pm \cos{\theta^*} + \gamma_B \frac{\beta^-}{\beta^+}  \left(  1 \mp \frac{\vec{v}_B \cdot \vec{u}_{\Omega_l^*}}{ 1+ \gamma_B^{-1}}     \right)     \right] \; ,
\end{equation}
where $\vec{u}_{\Omega_l^*}$ is a unit vector pointing in the $\Omega_l^*$ direction. The scalar product $\vec{v}_{\rm B} \cdot \vec{u}_{\Omega_l^*}$ can be written in the form
\begin{equation}
\label{eq:boost_v_scalar_u}
 \vec{v}_{\rm B} \cdot \vec{u}_{\Omega_l^*} = - \frac{\beta^-}{\beta^+} \cos \vartheta^* - \frac{q_T}{\gamma_{\rm B} M} \sin \vartheta^* \cos \varphi^* \; .
\end{equation}
Considering the relation between the lepton transverse momentum $l_T^\pm$ and rapidity $\eta^\pm$ with the remaining relevant variables,
\begin{equation}
 l_T^\pm \equiv \lvert \vec{l}_T^\pm  \rvert = \left(l^\pm \right)^2 - \left( l^{z,\pm} \right)^2 
\; \; , \; \; 
 \eta^\pm = \textrm{arctanh}\frac{l^{z,\pm}}{l^\pm}
\, ,
\end{equation}
we have all the necessary ingredients needed to impose the kinematical cuts set by the LHCb collaboration~\cite{LHCb:2012fja}, {\it i.e.}
\begin{equation}
\label{eq:LHCb_cuts}
 2 < \eta^\pm < 4.5
 \; , \quad
 l^\pm > 10 \mbox{ GeV}
 \; , \quad
 \left\{
  \begin{array}{l}
   l^\pm_T > 3 \mbox{ GeV} \hspace{0.55cm} \mbox{if} \hspace{0.35cm} M \le 40 \mbox{ GeV} \\
   l^\pm_T > 15 \mbox{ GeV} \hspace{0.35cm} \mbox{if} \hspace{0.35cm} M > 40 \mbox{ GeV}
  \end{array}
 \right.
\end{equation}
with the dilepton invariant mass in the range $ 5.5 \mbox{ GeV} < M < 120 \mbox{ GeV}$.
For comparison, we give predictions also in the ATLAS kinematics~\cite{Piccaro:2012kzm,Aad:2014qja}, which, however, are constrained to more central rapidity ranges than the  LHCb experiment. In the ATLAS configuration, we have
\begin{equation}
\lvert \eta^\pm  \rvert < 2.4 
 \; \; , \; \;
l_T^\pm > 6 \text{ GeV}  
 \; \; , \; \;
l_T^+ > 9 \text{ GeV or }  l_T^- > 9 \text{ GeV} \, ,
\end{equation} 
with the dilepton invariant mass in the range $ 12 \mbox{ GeV} < M < 66 \mbox{ GeV}$. We expect our formalism to be more accurate when describing the LHCb data than the ATLAS data since it corresponds to a more forward kinematics. 

We have used \textsc{Fortran} for our numerical analysis, in particular the {\tt Vegas} integrator~\cite{Lepage:1977sw} as implemented in the {\tt Cuba} library~\cite{Cuba:2005,ConcCuba:2015} and specific {\tt Cernlib} routines~\cite{cernlib}.
A two-loop running coupling with $\alpha_s\left(M_Z\right)=0.11707$ and five active quark flavors was also chosen.
The NLO~{\tt Mmht}~2014 sets~\cite{Harland-Lang:2014zoa} were used, as provided by the {\tt Lhapdf} Interface 6.2.1~\cite{Buckley:2014ana}, to calculate the collinear quark parton distribution functions. In the calculation of the unintegrated gluon density, the uncertainty stemming from the numerical multidimensional integration when combining Eqs.~(\ref{eq:dsigma}) and~(\ref{eq:sigma}) with Eq.~(\ref{eq:W-HEF}) or Eq.~(\ref{eq:W-CDA}) was steadily held below 0.5\%. The error bands of all the presented results were calculated by varying the factorization scale\footnote{In Ref.~\cite{Brzeminski:2016lwh} the factorization scale was set equal to the transverse mass of the exchanged boson, $\mu_F = M_T \equiv \sqrt{M^2 + q_T^2}$. We checked that the effect of using this choice with respect to ours is negligible.} in the range $M/4 \le \mu_F \le 4M$, while the renormalization scale $\mu_R$ was fixed to $M$.

\begin{figure}
\centering

 \includegraphics[scale=0.8,clip]{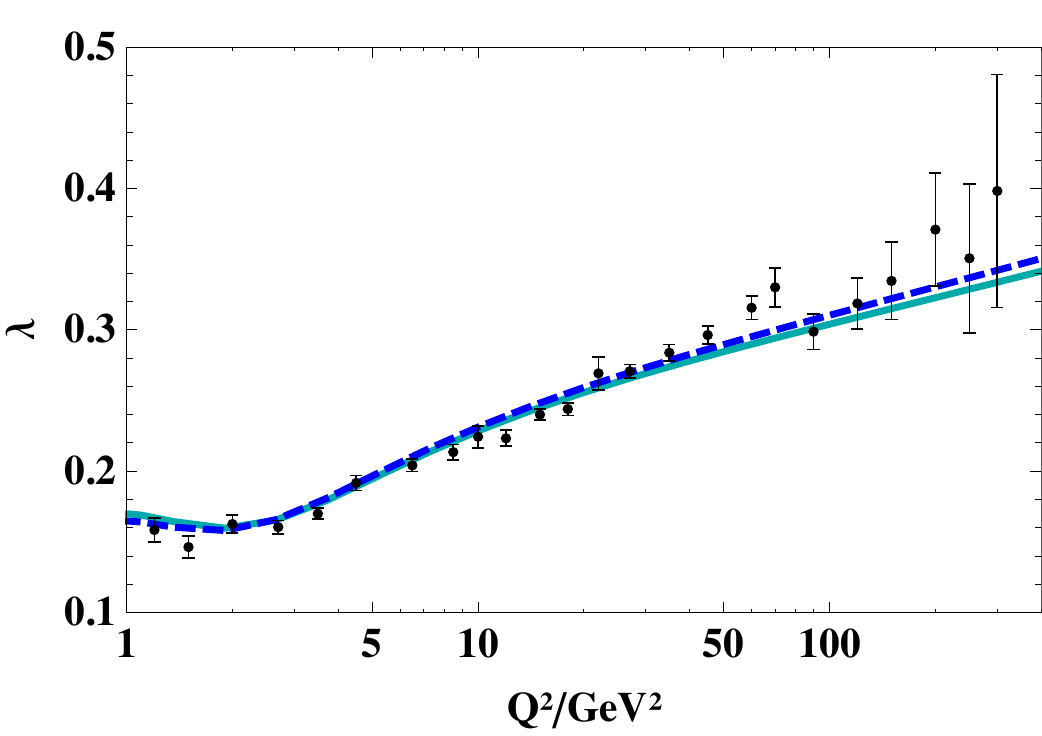}
 \text{a) Fit of the effective intercept $\lambda$ in $F_2 \simeq x^{- \lambda (Q^2)}$ for small $x$ HERA data.}
 \vspace{0.75cm}
 
 \includegraphics[scale=0.60,clip]{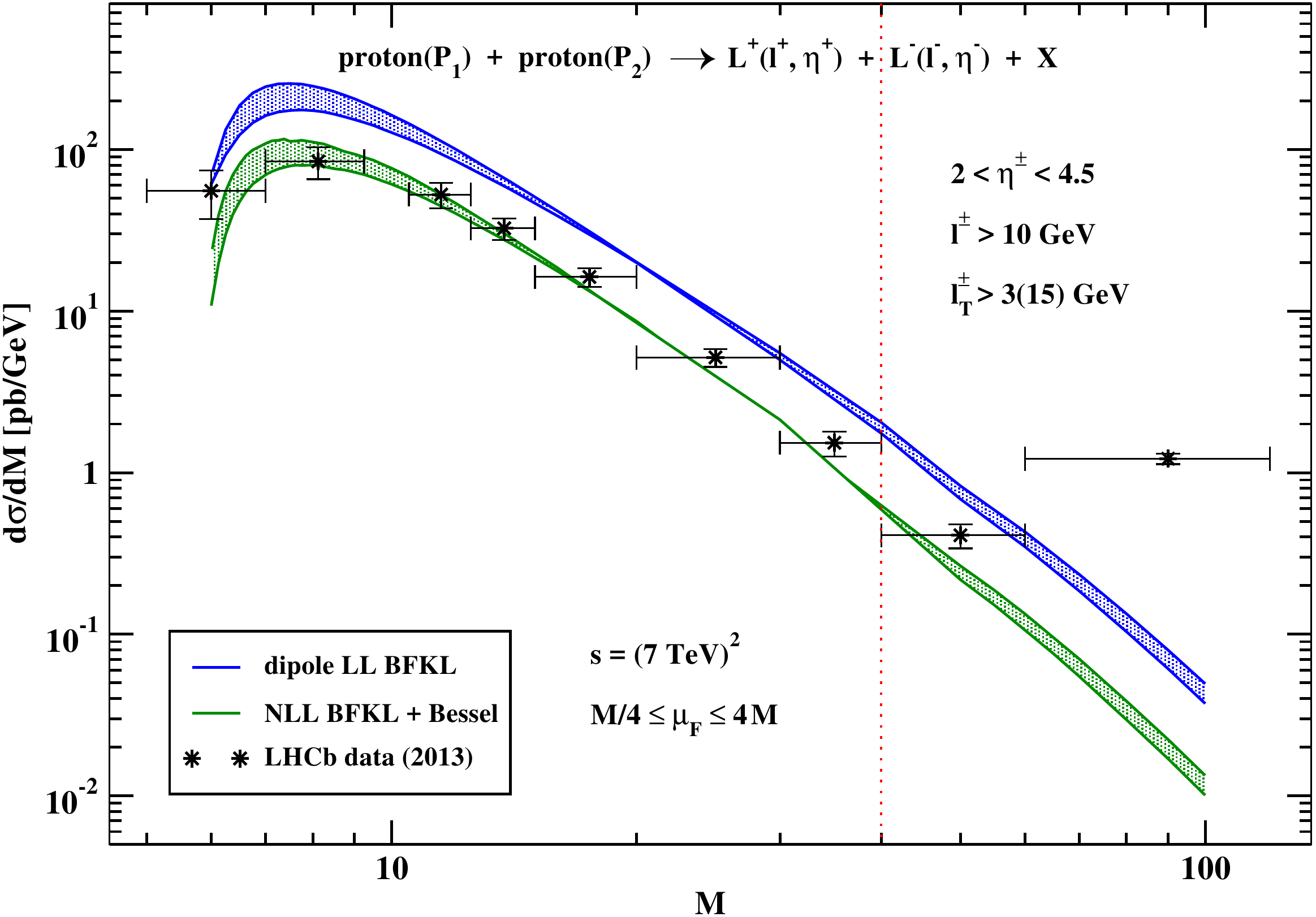}
 \text{b) $M$ dependence of the Drell--Yan differential cross section.}

\caption{Comparing experimental data with higher-order BFKL predictions for two different colliders. First (top panel, plot from Ref.~\cite{HERAfitH_SV_S}), the fit of the effective intercept $\lambda$ of $F_2$ to HERA data~\cite{Aaron:2009aa}. Solid and dashed lines refer to the LO photon impact factor and a kinematically improved one, respectively. Then (bottom panel), the Drell--Yan differential cross section is given as a function of the dilepton invariant mass $M$. The NLO  BFKL prediction with collinear corrections is compared with a LO dipole model and with LHCb data~\cite{LHCb:2012fja}. Uncertainty bands account for changes in the factorization scale in the range $M/4 \le \mu_F \le 4M$.}
\label{fig:fDY-NLO-LHCb}
\end{figure}

We present our results in Fig.~\ref{fig:fDY-NLO-LHCb} where we show how, in the lower plot, the outcome of our calculations is very close to the LHCb data points in the full range of $M$ values. In the same plot we also draw the line corresponding to the calculation in the dipole model approach with LO BFKL evolution and fixed running coupling, which lies well above the experiment results. For the sake of showing the universality of the model here presented, in the upper plot within the same figure, we reproduce the description given in~\cite{HERAfitH_SV_S} for the $Q^2$ dependence of the energy growth of the $F_2$ HERA structure function of the proton at small values of Bjorken $x$ when expressed in the form  $F_2 \simeq x^{-\lambda (Q^2)}$. Since, for simplicity, we do not include $Z$-boson production diagrams we lie slightly below the data for larger values of the DY invariant mass $M$. 

\begin{figure}
\centering

 \hspace{-0.90cm} \hspace{0.6cm}
 \includegraphics[scale=0.50,clip]{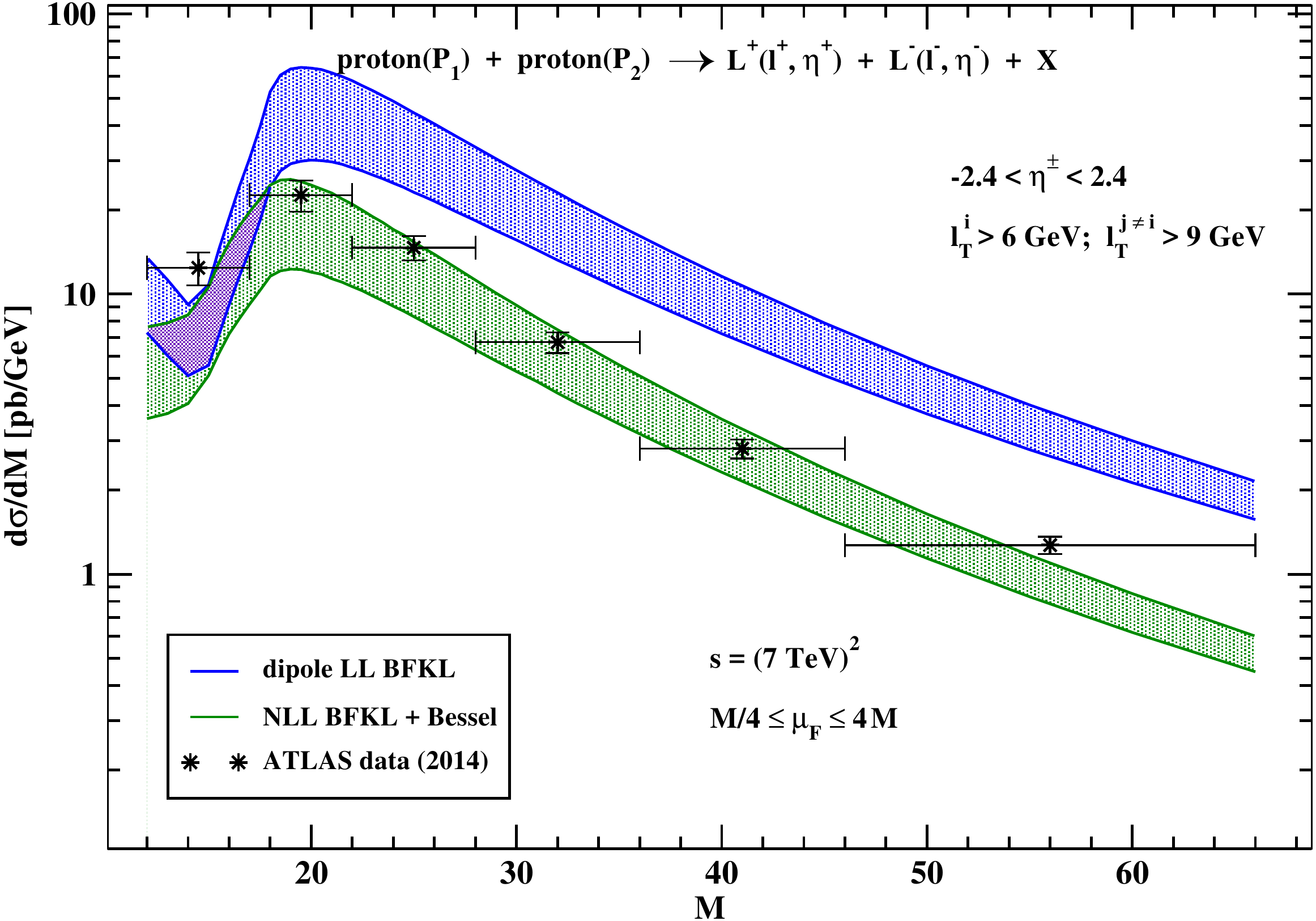}
\caption{Matching ATLAS data~\cite{Piccaro:2012kzm,Aad:2014qja} with higher-order BFKL predictions. Uncertainty bands are given as the effect of allowing the factorization scale to be in the range $M/4 \le \mu_F \le 4M$.}
\label{fig:fDY-NLO-ATLAS}
\end{figure}

The equivalent comparison to ATLAS data is presented in Fig.~\ref{fig:fDY-NLO-ATLAS}, which, as we have already mentioned, does not allow for a very forward production of the DY pair. Nevertheless, we obtain a good description of the data, even though the uncertainty band associated to changes in the factorization scale is rather large. Lower values of this scale seem to be preferred by the data as extracted from both experiments. 

\begin{figure}
\centering

 \includegraphics[scale=0.50,clip]{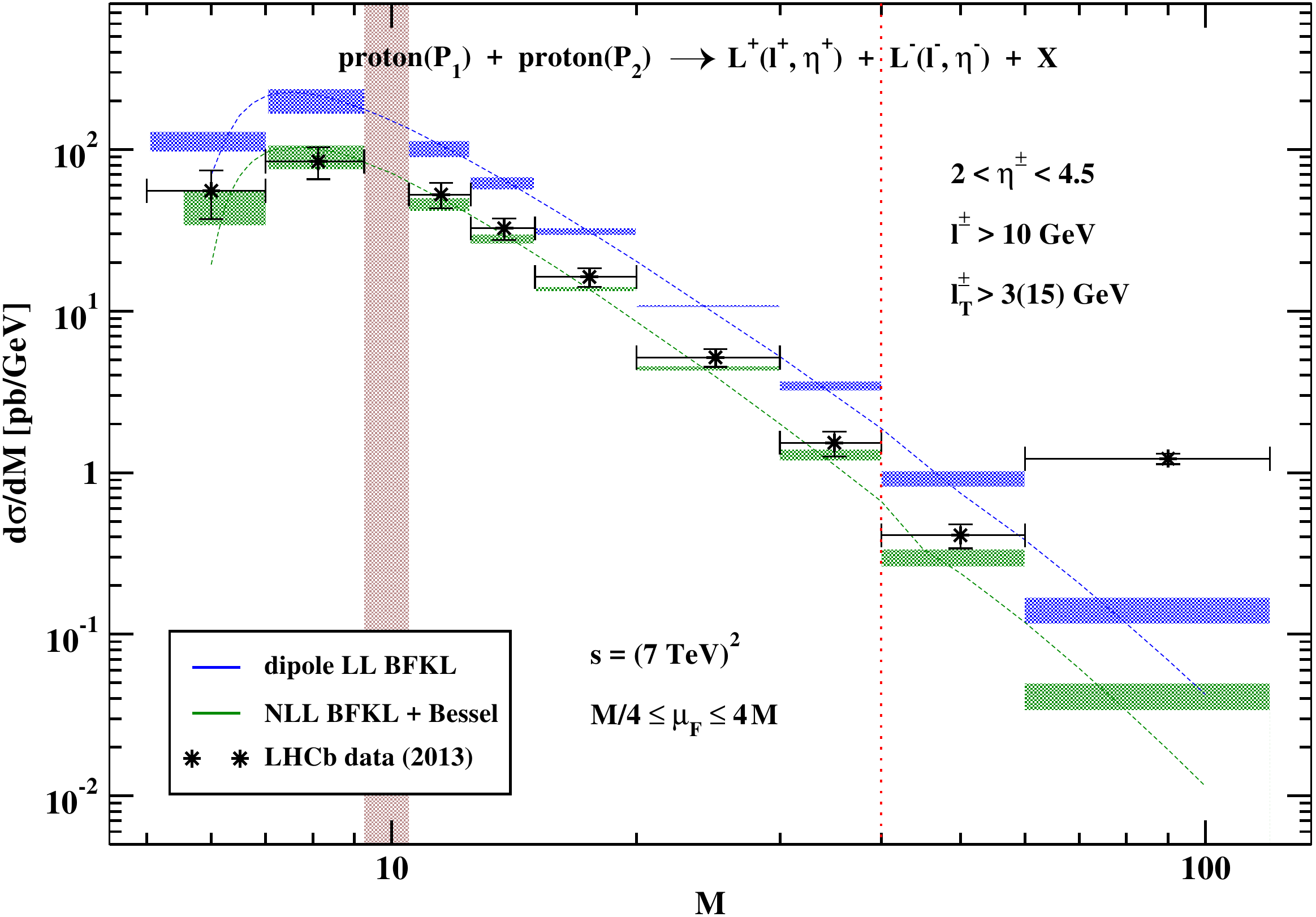}

\caption{Differential Drell--Yan cross section averaged over bins of the dilepton invariant mass $M$. The NLO BFKL with collinear corrections prediction is compared with a LO dipole model calculation and with LHCb data~\cite{LHCb:2012fja}. Uncertainty bands account for changes in the factorization scale in the range $M/4 \le \mu_F \le 4M$. Dashed lines show the $M$ dependence with $\mu_F = M$.}
\label{fig:fDY-NLO-LHCb-bin}
\end{figure}

Before presenting our Conclusions let us show the comparison of the  LHCb data with the cross section averaged on dilepton invariant mass bins. These provides a somehow more fair matching with the experimental presentation of the results. This can be seen in Fig.~\ref{fig:fDY-NLO-LHCb-bin}.

\section{Conclusions \& Outlook}

Building up on previous work on forward production of Drell--Yan pairs at the LHC in~\cite{Brzeminski:2016lwh},  we propose to use the BFKL formalism at next-to-leading order with collinear corrections to describe the LHCb and ATLAS data. We make use of the idea of high energy factorization and show that the same unintegrated gluon density as obtained from a fit of HERA data at small values of Bjorken $x$ provides a good description of the LHC data. This is an encouraging result from the point of view of the BFKL approach since this type of global description of different processes is expected from this framework. Nevertheless, the same data can be also  described by a fixed order calculation and this observable needs to be pushed experimentally further to really test different theoretical calculations. This also includes the description of the data in~\cite{Brzeminski:2016lwh} which makes use of saturation corrections. Future LHC data for Drell--Yan production in forward directions~\cite{N.Cartiglia:2015gve} will be very useful to gauge the need of high energy resummations in quantum field theory.

\section*{Acknowledgements}

We thank Grigorios Chachamis, Leszek Motyka and Tomasz Stebel for fruitful discussions and encouragement. 
ASV thanks Prof. Bryan Webber and the Cavendish Theory Group, at the University of Cambridge, for the warm hospitality during the summer of 2018, when this work was concluded. 
The research here presented has been supported by the Spanish Research Agency (Agencia Estatal de Investigaci\'on) through the grant IFT Centro de Excelencia Severo Ochoa SEV-2016-0597. ASV acknowledges further support from the Spanish Government grants FPA2015-65480-P, FPA2016-78022-P. DGG is supported with a fellowship of the international programme `La Caixa-Severo Ochoa'. FGC acknowledges support from the Italian Foundation ``Angelo della Riccia''.

\end{document}